\documentclass[10pt]{article}
\usepackage{algorithm2e}
\usepackage{graphicx}
\usepackage{subfig}
\usepackage{amsfonts}
\usepackage{amsmath}
\usepackage{array}
\usepackage{placeins}
\usepackage{color,soul}
\usepackage[noadjust]{cite}

\title{\LARGE \bf
A Driver-in-the Loop Fuel Economic Control Strategy for Connected Vehicles in Urban Roads
}

\author{Baisravan HomChaudhuri,$^{*1}$ and Pierluigi Pisu$^{2}$ % <-this % stops a space
\thanks{$^{1}$Baisravan HomChaudhuri is a Postdoctoral Fellow of Automotive Engineering,
       Clemson University, Greenville, SC 29607, USA
        {\tt\small baisravan.hc@gmail.com}}%
\thanks{$^{2}$Pierluigi Pisu is with the Department of Automotive Engineering, Clemson University,
        Greenville, SC 29607, USA
        {\tt\small pisup@clemson.edu}}%
}

\begin{document}

\maketitle
\thispagestyle{empty}
\pagestyle{empty}

%%%%%%%%%%%%%%%%%%%%%%%%%%%%%%%%%%%%%%%%%%%%%%%%%%%%%%%%%%%%%%%%%%%%%%%%%%%%%%%%
\begin{abstract}
In this paper, we focus on developing driver-in-the loop fuel economic control strategy for multiple connected vehicles. The control strategy is considered to work in a driver assistance framework where the controller gives command to a driver to follow while considering the ability of the driver in following control commands. Our proposed method uses vehicle-to-vehicle (V2V) communication, exploits traffic lights' Signal Phase and Timing (SPAT) information, models driver error injection with Markov chain, and employs scenario tree based stochastic model predictive control to improve vehicle fuel economy and traffic mobility. The proposed strategy is decentralized in nature as every vehicle evaluates its own strategy using only local information. Simulation results show the effect of consideration of driver error injection when synthesizing fuel economic controllers in a driver assistance fashion.
\end{abstract}
%%%%%%%%%%%%%%%%%%%%%%%%%%%%%%%%%%%%%%%%%%%%%%%%%%%%%%%%%%%%%%%%
%\keywords{Connected Vehicles, Stochastic Model Predictive Control, Driver-In-the Loop Control, Semi-Autonomous Systems}
%%%%%%%%%%%%%%%%%%%%%%%%%%%%%%%%%%%%%%%%%%%%%%%%%%%%%%%%%%%%%%%%%%%%%%%%%%%%%%%%
\section{INTRODUCTION}
Connected vehicle (CV) technology is considered to be a multi-faceted solution to the current issues of the transportation system as pointed out by the 
Intelligent Transportation Systems Joint Program Office (ITS-JPO). The CV technology proposed by the ITS-JPO allows vehicles to communicate with other vehicles (V2V communication) and with transportation infrastructure (vehicle to infrastructure communication (V2I)) via wireless communications. For ITS-JPO, the major areas of concern are safety, mobility, and the environmental impact of the vehicles.  According to the ITS-JPO website, traffic fatalities are more than $30$ thousand every year and traffic congestion costs around $87.2$ billion to the U.S. economy, with $4.2$ billion hours and $2.8$ billion gallons of fuel spent sitting in traffic \cite{2010cong}. Moreover, deterioration of traffic mobility affects the environment since vehicles that are stationary, idling, and traveling in a stop-and-go pattern emit more greenhouse gases than those traveling in free-flow conditions \cite{2010cong}.
Although the primary aim of CV systems is to improve safety (V2V and V2I communication addresses 79\% of all vehicle crashes \cite{najm2010frequency}), the V2V and V2I communication can be utilized to improve many different aspects of the transportation system such as improvement of fuel economy and traffic mobility, which in turn would improve vehicle emissions \cite{2010cong}. 

%The primary aim of the envisioned CV technology is to improve safety of the transportation system since it has been reported that.
% In the CV environment, vehicles can be considered to share their position, velocity and acceleration information to its neighboring vehicles while traffic road side units (RSUs) can share SPAT information of the traffic signals to the CVs. 
%In the connected vehicle environment, information such as position, velocity, and acceleration of a vehicle can be considered to be shared with other vehicles while traffic information, such as traffic light timing and road grade information, can also be made available to the communicating vehicles. This added information, apart from improving safety, is very useful in improving other aspects such as developing fuel economic control strategies for the vehicles.

In recent years, a lot of research in intelligent transportation system and automotive engineering has focused on developing more fuel efficient vehicles. The fuel efficiency of a vehicle depends on a number of factors such as vehicle aerodynamic drag, vehicle engine characteristics and powertrain system, and weather and road conditions. 
%Hybrid electric vehicles (HEVs) or plug-in electric vehicles (PEVs) have been developed to improve fuel efficiency, especially for urban road conditions, as well as for the reduction of environmental footprints of the vehicles. Despite the advantages of HEVs and PEVs, they only constitute a very small fraction of the total number of vehicles which necessitates the development of fuel efficient conventional vehicles.
Apart from them, driving behavior of vehicles have been seen to influence fuel efficiency by a good margin \cite{van2004driving}. It has been observed, that fuel efficiency of a vehicle improves when they move at a constant cruising velocity and that is why most fuel economic control strategies in the literature tries to minimize vehicle acceleration and braking \cite{gilbert1976vehicle, chang2005vehicle,hooker1988optimal,kirschbaum2002determination,hellstrom2010design,kamal2010ecological, kamal2013model,van2004driving}. 

Developing fuel efficient control strategies for urban roads is difficult because of traffic lights at regular intervals. However, reduction of red light idling have shown to improve vehicle's fuel efficiency \cite{asadi2011predictive,mahler2012reducing}. That is why, in recent years, authors  \cite{asadi2011predictive, mahler2012reducing, de2013eco, mandava2009arterial, rakha2011eco, kamalanathsharma2013multi} have developed control strategies that utilize traffic light timing information to avoid red light idling. 
Authors \cite{mandava2009arterial,asadi2011predictive,ozatay2013analytical} have developed methods to optimally control vehicle acceleration and velocities while moving through multiple traffic signals. A probabilistic decision making algorithm using model predictive control is shown in \cite{mahler2012reducing} that addresses noise in SPAT information. Fuel optimal control strategy using data driven fuel consumption model is shown in \cite{rakha2011eco}, while a multi-stage dynamic programming based control strategy is provided by the authors in \cite{kamalanathsharma2013multi}. Field testing of fuel optimal control strategies using SPAT information with real vehicles is shown by the authors in \cite{xia2012field}. When traffic light timings are not known, researchers in \cite{ozataybayesian} have shown a stochastic online estimation method to estimate the parameters of traffic lights. Most of the above mentioned works are designed for single vehicle systems which for urban roads is a limiting condition. Although some literature \cite{kamal2013model} have considered two vehicle scenarios, the velocity and acceleration prediction strategy for the preceding vehicle might not be feasible for congested roads.
%Most common control strategies used by vehicles in multi-vehicle scenarios are the car following models, such as the Gipp's car following model \cite{gipps1981behavioural}, intelligent driver model \cite{treiber2000congested} and enhanced intelligent driver model \cite{kesting2010enhanced}, but they do not focus on improvement of fuel economy.
Also, all the above works consider full autonomous vehicles where the evaluated control commands are exactly executed, which is a hard assumption when human drivers are present. 
Full autonomous driving in urban roads is a complex task \cite{bengler2014three} and a percentage of the population can always be expected to drive by choice.  
This demands for development of control methods that work together with the human driver and addresses their imperfection while following given commands.

Driver assistance systems have gained a lot of popularity in the recent years and they primarily focus on vehicle safety (such as giving warning signals). These assistance systems work with the driver by either giving them commands to follow, such as maintaining a velocity, or provide active control for a short-while when required. The common driver assistance systems include brake assistance system, driver status monitoring, blind spot warning, lane departure warning and lane keeping assistance system, speed control, and night vision system \cite{hummel2011advanced}. These systems have evolved over time and some exploit the recent advances in CV technology \cite{shladover2012literature,bengler2014three,piao2008advanced}. 
Some recent works on driver assistant systems that focus on vehicle fuel economy are available in \cite{wu2011fuel,vagg2013development,gilman2015personalised, guan2012fuel}. Authors \cite{wu2011fuel} have provided optimal velocity and acceleration information to the driver by minimizing an approximate fuel consumption model. A driver training and analysis system is developed in \cite{vagg2013development} where the drivers are taught to reduce sharp acceleration and braking. A context aware driver assistance system is proposed in \cite{gilman2015personalised} that considers driver, environmental, and in-vehicle conditions for its decision making. Authors in \cite{guan2012fuel} have presented a driver assistance system that focus of fuel economy using data driven and manufacturer independent model.
Most of the above mentioned works on driver assistance systems do not consider the driver behavior or his/her capabilities in following commands from the assistance system.  Apart from that, most of the work in the literature focuses on developing fuel economic control strategies for a single vehicle without considering the impact of a vehicle's driving behavior on another.

%Driver assistance system, although popular, generally does not consider driver behavior or their capabilities (if they can follow the instructions), especially not online and when focusing on fuel economy.
Some previous research on semi-autonomous control of vehicles are available in \cite{di2014stochastic,gray2013stochastic, shia2014semiautonomous, gray2012semi}. Authors \cite{di2014stochastic} have used stochastic model predictive control (SMPC) for energy management in hybrid electric vehicles (HEVs) where Markov chain model is used to model driver future power request. Authors in \cite{gray2013stochastic, shia2014semiautonomous, gray2012semi} have developed a controller that only takes over the vehicle when it predicts some danger, e.g., vehicle entering wrong lane, and have modeled the human driver as a feedback controller.
%In automotive engineer, stochastic model predictive control is used in \cite{di2014stochastic} and \cite{gray2013stochastic} for energy management in hybrid electric vehicles (HEVs) and driver-in the loop vehicle lateral control respectively. In the former paper, the authors use Markov model based driver prower request and use scenario based stochastic model predictive control for energy management in HEVs. In \cite{gray2013stochastic}, the authors develop active control based strategy for vehicle lane keeping in presence of human drivers.
In CV systems and urban road conditions, longitudinal vehicle control in a driver assistance framework that focus on fuel economy and system mobility while considering driver behavior has not been addressed before. Thus, in this paper, we develop a driver-in-the loop fuel economic control strategy for multiple CVs in urban road conditions that also focus on the improvement of system mobility by reducing red light idling using SPAT information. Since the controller design is driver specific, the assistance system can be considered to be a \emph{customized/personalized driver assistance system}. 
The developed strategy is decentralized in nature as every vehicle can develop its own  control using only local V2V and V2I information. First we model the driver error with a Markov chain model and then employ a scenario based stochastic model predictive control strategy \cite{bernardini2009scenario, di2014stochastic, grosso2013assessment, leidereiter2014quadrature,heitsch2009scenario,bernardini2012stabilizing} for evaluating the optimal control policy in presence of a human driver. 

In some of our previous works \cite{HomChaudhuri2015, homchaudhuri2016hierarchical,lin2015fuel, HomChaudhuri2016}, we have developed model predictive control based fuel economic control strategies for a group of conventional and hybrid electric vehicles \cite{homchaudhuri2016hierarchical} in urban roads. In this paper, we extend those works by developing the control strategy as a driver assistance system while considering the driver behavior. The paper contributions can be listed as: (i) a longitudinal vehicle control strategy is developed in a driver assistance framework considering the driver behavior, especially the imperfection associated with the driver when following the instructions from the assistance system; (ii) consideration of computational tractability by employing sampling based scenario tree generation and use of discounted cost for the SMPC; and (iii) consideration of multiple CVs and an urban environment (presence of traffic lights) for fuel economic controller development. While improving fuel efficiency of vehicles, we also improve system mobility which in turn improves vehicle environmental impact \cite{2010cong, homchaudhuri2016hierarchical}. We have compared our proposed method with a passive driver assistance system (does not consider human error injection) and with two popular optimal control methods for stochastic systems: ($i$) certainty equivalence control \cite{stengel1986stochastic} , and ($ii$) frozen time model predictive control \cite{di2014stochastic,bernardini2012stabilizing}. 

The paper is organized as follows: in Section \ref{problem}, we mathematically introduce the problem and in Section \ref{approach}, we describe the stochastic optimal control strategy that uses V2I SPAT information and V2V neighboring vehicle information. Finally, in Sections \ref{simulation} and \ref{conclusion}, we provide the simulation results and the conclusion and the future potentials of this work.
%%%%%%%%%%%%%%%%%%%%%%%%%%%%%%%%%%%%%%%%%%%%%%%%%%%%%%%%%%%%%%%%%%%%%%%%%%%%%%%%%
\section{PROBLEM DESCRIPTION}\label{problem}
\begin{figure}[h]
%\graphicspath{{./fig/}}
\centering
\includegraphics[width=3.2in]{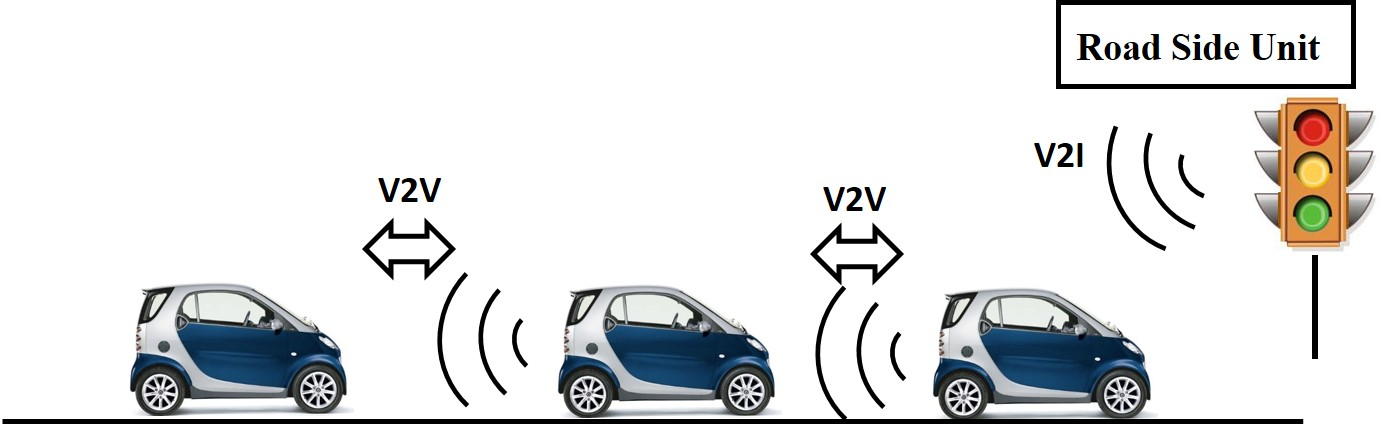}
\caption{Schematic of the problem}
\label{schematic}
\end{figure}

\begin{figure}[h]
%\graphicspath{{./fig/}}
\centering
\includegraphics[width=3.4in]{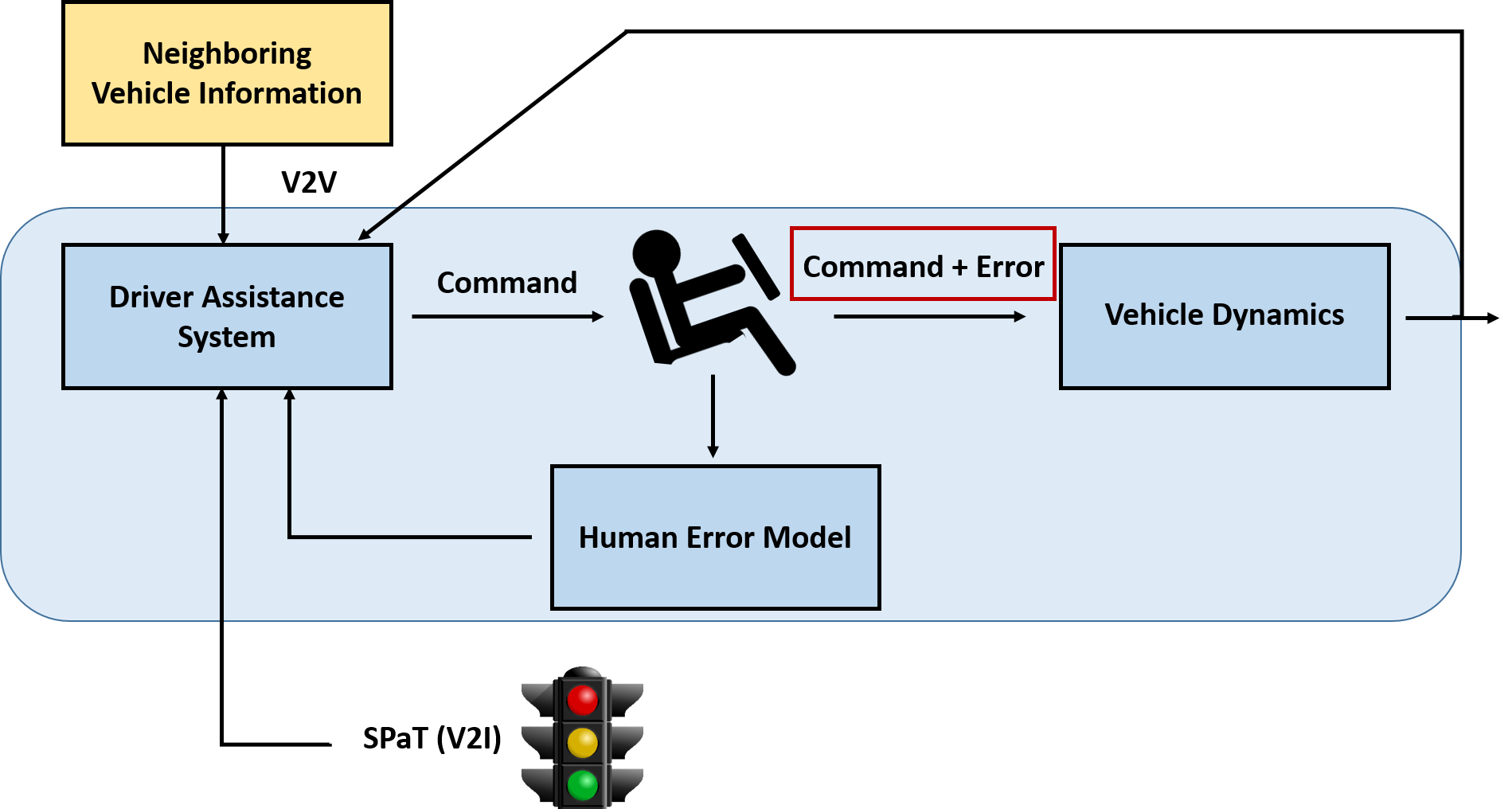}
\caption{Personalized driver assistance system}
\label{schematic_driver}
\end{figure}

We focus on a group of CVs in urban road conditions that consist of traffic lights at regular intervals. In this CV scenario, wireless V2V communication allow vehicles to share their position and velocity information with  other CVs. In every vehicle, the driver assistance system is considered to provide velocity command to the driver which the driver needs to follow to improve its fuel economy. For human drivers, following these commands exactly at every time instant is difficult and their actions would introduce random errors which would be driver specific. Fig. \ref{schematic} shows a schematic of the mentioned scenario and Fig. \ref{schematic_driver} shows the driver assistant system in every vehicle. 
%In this section, we present the mathematical description of the problem while in the next section we discuss in detail the methods employed in modeling and handling the random error introduced by the human drivers.

%%%%%%%%%%%%%%%%%%%%%%%%%%%%%%%%
\subsection{System Description}
The discrete time longitudinal dynamics of any vehicle $i$ with time step $\Delta t$ is given by \cite{kamal2013model}:
%\begin{align} \label{state_space1}
%\begin{aligned}
%&\dot{x_i} = f_i(x_i,u_i^f)\\
%&f_i(x_i,u_i^f) = \left [ \begin{array}{c}
%v_i\\
%-\frac{1}{2M_h^i}C_D\rho_aA_v^i v_i^2 -\mu g -g\theta +u_i^f 
%\end{array} \right] \\
%&u_i^f = u_i + \omega_i
%\end{aligned}
%\end{align}
\begin{align} \label{state_space1}
\begin{aligned}
%x_i(k+1) &= f_i(x_i(k),u_i^f(k))\\
{x_i}(k+1) &=x_i(k) + \Delta t f_i(x_i(k),u_i^f(k))\\
& \hspace*{-1cm}\text{with } \\
f_i(x_i(k),&u_i^f(k)) \\
&= \left [ \begin{array}{l}
v_i(k)\\
-\frac{1}{2M_h^i}C_D\rho_aA_v^i v_i^2(k) -\mu g - g\theta +u_i^f(k) 
\end{array} \right] \\
u_i^f(k) &= u_i(k) + \omega_i(k)
\end{aligned}
\end{align}
Here $x_i(k) \in \mathbb{R}^{n_x}$ and $u_i^f(k) \in \mathcal{U} \subseteq \mathbb{R}^{n_u}$, where $n_x = 2$ and $n_u=1$. The vehicle state is $x_i(k) = [s_i(k) \:\: v_i(k)]^T$, where $s_i(k)$ is the position of a vehicle while $v_i(k)$ is its velocity at time instant $k$. In the above equation, $M_h^i$ is the mass of the vehicle $i$, and $A_v^i$ is its  frontal area. The terms $C_D$, $\rho_a$, $\theta$ and $\mu$ are the drag coefficient, air density, the road gradient, and the rolling friction coefficient respectively.  It is assumed that road gradients are mild, so $sin(\theta) = \theta$ and $cos(\theta) = 1$. The effective control strategy, $u_i^f(k)$, of a vehicle $i$ is its braking or traction force per its unit mass which is considered to be a sum of the control, $u_i(k)$, suggested by the assistance system and the random error, $\omega_i(k) \in \mathcal{W}$, injected by the human driver. The constraint sets $\mathcal{U}$ and $\mathcal{W}$ are considered to be compact.
We can consider that the assistance system provides velocity information to the driver to follow so that at $k$, it would ask the driver to reach velocity $v_i(k+1) = v_i(k) + \Delta t u_i(k)$ when the vehicle velocity is $v_i(k)$.

%In the above equation, $M_h^i$, $C_D$, $\rho_a$, $A_v^i$, $\mu$ and $\theta$ are the mass of the vehicle, drag coefficient, the air density, the frontal area of the vehicle, the rolling friction coefficient and the road gradient respectively.

As mentioned before, the fuel consumption in a vehicle depends on a number of factors such as torque, engine speed, and gear ratio \cite{kamal2010ecological}. Many papers  approximate the cost of fuel consumption as a function of velocity and acceleration of the vehicle. Following \cite{kamal2013model}, we represent the rate of fuel consumption (ml/s) by:
%\begin{subequations}\label{fuel_cons}
%\begin{align}
%\dot{Fuel}_i(t) &= (1-\zeta)(\dot{f}_{cruise}^i(t) + \dot{f}_{accel}^i(t)) + \zeta F_d^i(t) \label{fuel1}\\
% \dot{f}_{cruise}^i(t) &= b_0 + b_1 v_i(t) + b_2v_i(t)^2 + b_3v_i(t)^3 \label{fuel2}\\
% \dot{f}_{accel}^i(t) &= \hat{a}_i(c_0 + c_1 v_i(t) + c_2v_i(t)^2) \label{fuel3}\\
% \hat{a}_i &= -\frac{1}{2M_h^i}C_D\rho_aA_v^i v_i(t)^2 -\mu g  + u_i^f \label{fuel4}\\
% \zeta &= \begin{cases}
% 1 & \text{if $v_i(t) = 0$ or $u_i <0$} \notag \\ 
% 0 & \text{otherwise} \notag
% \end{cases}
% \end{align}
%\end{subequations}
\begin{subequations}\label{fuel_cons}
\begin{align}
\dot{Fuel}_i &= (1-\zeta)(\dot{f}_{cruise}^i + \dot{f}_{accel}^i) + \zeta F_d^i \label{fuel1}\\
 \dot{f}_{cruise}^i &= b_0 + b_1 v_i + b_2v_i^2 + b_3v_i^3 \label{fuel2}\\
 \dot{f}_{accel}^i &= \hat{a}_i(c_0 + c_1 v_i + c_2v_i^2) \label{fuel3}\\
 \hat{a}_i &= -\frac{1}{2M_h^i}C_D\rho_aA_v^i v_i^2 -\mu g  + u_i^f \label{fuel4}\\
 \zeta &= \begin{cases}
 1 & \text{if $v_i = 0$ or $u_i^f <0$} \notag \\ 
 0 & \text{otherwise} \notag
 \end{cases}
 \end{align}
\end{subequations}
In Eq. (\ref{fuel_cons}), $\dot{f}_{cruise}^i$ is the rate of fuel consumed while cruising and $\dot{f}_{accel}^i $ is the rate of fuel consumed when accelerating. Apart from that, a constant fuel consumption $F_d^i$ is considered when red light idling and braking \cite{kamal2010ecological}. The binary term $\zeta$ is $1$ during braking or red light idling and $0$ otherwise. The constant terms $b_0$, $b_1$, $b_2$, $b_3$, $c_0$, $c_1$ and $c_2$ are specific to a vehicle and could be found to approximate the fuel map of a vehicle.
%The fuel consumed per unit time for any vehicle $i$ is given by Eq. (\ref{fuel1}).
%Apart from the fuel consumption due to motion, a constant fuel cost of $F_d^i$ (ml/s) is considered to be associated with braking and idling at a red light .
%%%%%%%%%%%%%%%%%%%%%%%%%%%%%%%%%%%%%%%%%%%%%%%%%%%%%%%%%%%%%%
\subsection{Optimal Control Problem}\label{veh_cost}
To maximize `miles per gallon' (mpg), we can minimize fuel consumption per unit distance. The fuel consumed per unit distance for a group of $n$ vehicles is given by $\sum_{i=1}^{n} \sum_k [(\dot{Fuel}_i(k) \Delta t)/(v_i(k)\Delta t )]$. 
%The optimal control problem for any vehicle $i$ is thus given by:
%is considered to be the measure of the fuel efficiency
%Considering a time horizon given by $T$ seconds, the cost function associated with each vehicle $i$ at each time instant $k$ is given by \cite{kamal2010ecological}:
%\begin{subequations}\label{cost_old1}
%\begin{align}
%\argmin \limits_{u_i} & \quad \mathbb{E} \left [\sum_{k}\frac{\dot{Fuel}_i(k)\Delta t}{v_i(k)\Delta t}\right]  \label{cost_old}\\
%v_{min}&\leq v_i(k)\leq v_{max}, \forall k \label{cons1}\\
%u_{min}^i&\leq u_i^f(k)\leq u_{max}^i,  \forall k \label{cons2}
%\end{align}
%\end{subequations}
%$\mathbb{E}[\cdot]$ in the above equation is the expectation value. Since the human driver introduces the random error while trying to follow the command of the driver assistance system, the overall goal here is to minimize the expected value of the fuel consumption cost. The above problem maximizes the expected miles per gallon of every vehicle where the control variable is the vehicle's traction or braking force generated by its powertrain. In Eq. (\ref{cons1}), $v_{min}$ and $v_{max}$ are the minimum and maximum road speed limits and in Eq. (\ref{cons2}), $u_{min}^i$ and $u_{max}^i$ are the bounds on the effective control $u_i^f(k)$. 
%Apart the constraints shown in Eq. (\ref{cost1}), the system constraints given by Eq. (\ref{state_space1}) is also required to be satisfied.
Generally, this problem is solved as a receding horizon problem, with a time horizon $T$ secs, where the cost includes fuel consumption per unit distance, collision avoidance penalty, vehicle's desired velocity tracking, and the cost of applying control. The optimal control problem for every vehicle $i$ is given by \cite{kamal2010ecological}:
\begin{subequations}\label{cost1}
\begin{align}
&\hspace{1cm} \min \limits_{u_i} \:\: J_i(k) \\
J_i(k) &=  \mathbb{E} \bigg[\sum_{t=k}^{k+M-1} c_1 \frac{\dot{Fuel}_i(t)\Delta t}{v_i(t) \Delta t} + c_2^i(t) R_{ij}(t)^2  \notag\\
 &  + c_3 (v_i(t)  -v_{target}^i(k))^2 + c_4 u_i^f(t)^2  \bigg] \label{cost}\\
R_{ij}(t) &= S_0 + t_{hd}(v_i(t)-v_j(t)) + (s_i(t) - s_j(t)) \label{rij}\\
v_{min}&\leq v_i(t)\leq v_{max} \label{cons1}\\
u_{min}^i&\leq u_i^f(t)\leq u_{max}^i \label{cons2}
\end{align}
\end{subequations}
In the above equation, $M = (T/\Delta t)$ is the discrete time horizon and $\mathbb{E}[\cdot]$ is the expectation. Since the human driver introduces the random error while trying to follow the command of the driver assistance system, the overall goal here is to minimize the expected cost.
 In Eq. (\ref{cost}), the first term inside the expectation minimizes fuel consumed per unit distance, the second term minimizes the deviation from a desirable distance between vehicle $i$ and its preceding vehicle $j$, the third term tries to minimize the velocity deviation from its target velocity ($v_{target}^i(k)$), and the last term minimizes the control effort. The target or desired velocity of vehicle $i$ at time $k$ is given by $v_{target}^i(k)$ and is generally chosen as the road speed limit. The terms $S_0$ and $t_{hd}$ in Eq. (\ref{rij}) are  predefined critical distance and headway time respectively. In Eq. (\ref{cost}), $c_1$, $c_3$ and $c_4$ are constant weights while $c_2^i(t)$, similar to \cite{kamal2013model}, is chosen as a function of the relative distance, $(s_j(t) - s_i(t))$, so that it increases as the relative distance decreases and vice versa. 
% The above problem is needed to be solved considering the constraints in Eq. \ref{rij}, Eq. (\ref{cons1}), Eq. (\ref{cons2}), and the state dynamics equation Eq. (\ref{state_space1}) which includes the random driver error.

%%%%%%%%%%%%%%%%%%%%%%%%%%%%%%%%%%%%%%%%%%%%%%%%%%%%%%%%%%%%%%%%%%%%%%%%%%%%%%
\section{METHODOLOGY}\label{approach}
Our proposed methodology works in two phases: $(A):$ we evaluate the target velocity of a vehicle so that it avoids red light idling at the upcoming traffic signal, and $(B):$ a scenario tree based stochastic model predictive control strategy is proposed for each vehicle that aims at improving their performance. 
%%%%%%%%%%%%%%%%%%%%%%%%%%
\subsection{Target Velocity} \label{target_velocity}
We consider the SPAT information of only the upcoming traffic signal is known by each vehicle, via V2I communication. 
%To reduce the impact of this noise, a vehicle choose a shorter green light interval ($t_g$), than the actual interval ($t_g<t_g^{actual}$), so that every time the vehicle experiences a green light at the traffic signal in that shorter interval.
Rather than using speed limit (maximum allowable velocity) as the target velocity as in \cite{kamal2013model}, each vehicle evaluates its target velocity so that it can avoid red light idling at the upcoming traffic signal. The \emph{target velocity} ($v_{target}^i(k)$) is computed by each vehicle $i$ as \cite{asadi2011predictive, HomChaudhuri2015, HomChaudhuri2016}:
\begin{subequations} \label{v_tar1}
\begin{align}
&v_{target}^i(k) = 
	\begin{cases} \frac{d_{iq}(k)}{K_wt_{cycle}-t_g - k}  & \text{if light =red} \\
		v_{max}  & \text{if light =green and}  \:\: \\
		& \frac{d_{iq}(k)}{K_wt_{cycle} -k} \leq v_{max} \\
		\frac{d_{iq}(k)}{K_wt_{cycle} + t_r - k} & \text{if light =green and Otherwise} 
	\end{cases}  \label{v_tar}\\
& \text{light = }
     \begin{cases}
     	\text{red} & \text{if } 0 \leq \text{mod}(\frac{k}{t_{cycle}}) \leq t_r \\
     	\text{green} & \text{if  } t_r < \text{mod}(\frac{k}{t_{cycle}}) < t_{cycle}
     \end{cases}\\
& v_{min}\leq v_{target}^i(k) \leq v_{max} \label{cons_target} \\
& t_{cycle} = t_r + t_g \\
&  K_w>\frac{k}{ t_{cycle}} \label{k_kw}
\end{align}
\end{subequations}
Here $d_{iq}(k)$ is the distance between $s_i(k)$ (location of the $i^{th}$ vehicle) and the traffic signal $q$, $t_r$ and $t_g$ are the red and green light durations respectively so that the total cycle duration is $t_{cycle}$. $K_w$ is an integer describing the traffic light cycle number. The function $\text{mod}()$ is a modulo function which generates the residue of division $k$ by $t_{cycle}$. It can be seen from Eq. (\ref{v_tar}) that when traffic light status is green, the maximum allowable speed is chosen as the target velocity unless the constraint $\frac{d_{iq}(k)}{K_wt_{cycle}-k} \leq v_{max}$ is not satisfied. Violation of this constraint (when the signal status is green) means that the vehicle needs to break the speed limit to pass through the traffic light in the current green light window. In that case, the vehicle desires to pass through the traffic signal in the next green light window as shown in third case of Eq. (\ref{v_tar}). If no feasible velocity is obtained in the consecutive green light windows, the vehicle has to stop at the given traffic light signal. Eq. (\ref{k_kw}) shows the constraint on the traffic signal index number so that $K_w$ is increased by 1 at $k = K_wt_{cycle}$. 
%In short, the target velocity is a feasible velocity (given by Eq. (\ref{cons_target})) which is chosen to make a vehicle move past a traffic signal, $a$, through a \emph{green light window}.

%%%%%%%%%%%%%%%%%%%%%%%%%%%%%%%%%%%%%%%%%%%%%%%%%%%%%%%%%%%%%%%%%%%%%%%
\subsection{Stochastic Model Predictive Control}\label{mpc}

%Stochastic optimal control problems can be solved using stochastic dynamic programming \cite{bertsekas1995dynamic} but they are very computationally expensive. 
Stochastic model predictive control (SMPC) \cite{gray2013stochastic, di2014stochastic,bernardini2012stabilizing} is a popular method for solving constrained stochastic optimal control problems.
%Some other approaches to solving optimal control problems for stochastic systems that require less computational efforts are certainty equivalent optimal control \cite{stengel1986stochastic}, stochastic model predictive control \cite{gray2013stochastic, di2014stochastic,bernardini2012stabilizing}, and frozen time model predictive control \cite{di2014stochastic,bernardini2012stabilizing}. 
%In certainty equivalent optimal control problem, the future random error realizations are assumed to be known and often replaced by the expected error while in frozen time model predictive control, the future error realizations (over the horizon) are considered to be constant and equal to the initial error. Stochastic model predictive control has become a popular method for solving receding horizon control problems with system uncertainties as can be found in \cite{hooshmand2012stochastic, di2014stochastic,grosso2013assessment, gray2013stochastic}. Authors in \cite{hooshmand2012stochastic} and \cite{grosso2013assessment} have applied this method to micro-grids and water networks respectively.
Thus, after the target velocity of a vehicle $i$ is evaluated, the driver-in the loop fuel efficient control problem is solved using SMPC. Solving problem in Eq. (\ref{cost1}) is challenging because it involves computation of the expected cost and solving the problem over the finite horizon $T$ secs. 
%We discretize the error $\omega_i$ into a finite number of states $n_{\omega}$ so that the probability of $\omega_{i}=q$ is given by $p_q$, i.e., $\mathbb{P}(\omega_{i}(k)=q) = p_q$. 
For computational purposes, we discretize the error set $\mathcal{W}$ (hence $\mathcal{W}$ is a bounded countable set) into $|\mathcal{W}|$ parts ($|\mathcal{W}|$ is the cardinality of the set $\mathcal{W}$). The cardinality of the set $\mathcal{W}$ here dictates the trade-off between accuracy and computational requirement since large $|\mathcal{W}| $ would reduce discretization error but increase computational requirement. 

In the literature, human behavior has been modeled with Markov chains by many researchers \cite{pentland1999modeling,takano2008recognition,wang2009hmm,lam2014pomdp,gray2002markov,sycara2015abstraction,gray2013stochastic, di2014stochastic}. Authors in \cite{gray2002markov,sycara2015abstraction} have used Markov model for analytical representation of human cognitive process. Authors in \cite{pentland1999modeling} modeled human behavior as a set of dynamic models sequenced together by Markov chain. Authors in \cite{takano2008recognition,wang2009hmm} used hidden Markov model while authors in \cite{lam2014pomdp} used partially observable Markov decision process model for human behavior modeling. Thus error introduced by humans in a system can be well modeled as a Markov chain. Markov chain model of human error can further be justified from the fact that human actions are highly dependent on the current state of the system.

Thus, following previous works, we model the drivel error with a Markov chain with transition probability matrix $\mathcal{Q}_i \in \mathbb{R}^{|\mathcal{W}|\times |\mathcal{W}|}$ where its elements $\mathcal{Q}_i(a,b)$ ($a^{th}$ row and $b^{th}$ column) are given by:
\begin{equation}\label{transition_prob}
\mathcal{Q}_i(a,b) = \mathbb{P}(\omega_i(k+1) = b| \omega_i(k) = a)
\end{equation}
 Here $\mathcal{Q}_i(a,b)$ gives the probability of state transition from state $a$ to $b$. The transition probability matrix would be driver specific and it can be modeled by the methods shown in \cite{di2014stochastic}. 
% The error $\omega_i(k)$ is the difference between the control command $u_i(k)$ from the assistance system and the actual driver input $u_i^f(k)$. Thus at current instant $k$, it is measurable (the difference can be computed easily) but its future values are unknown and random. 
% Error discretization helps in  computational purposes (large $|\mathcal{W}| $ would reduce discretization error but increase computational requirement) and Markov chain model helps in modeling the stochastic behavior of the human drivers while tracking commands from the driver assistance system. 
It can be seen from Eq. (\ref{state_space1}) that different values of $\omega_i(\cdot)$ represents different state dynamics and over a finite horizon, many different state trajectories are possible depending on the future values of $\omega_i(\cdot)$. It may be noted here that no assumption on the probability distribution of the error is made.

\begin{figure}[h]
%\graphicspath{{./fig/}}
\centering
\includegraphics[width=3in]{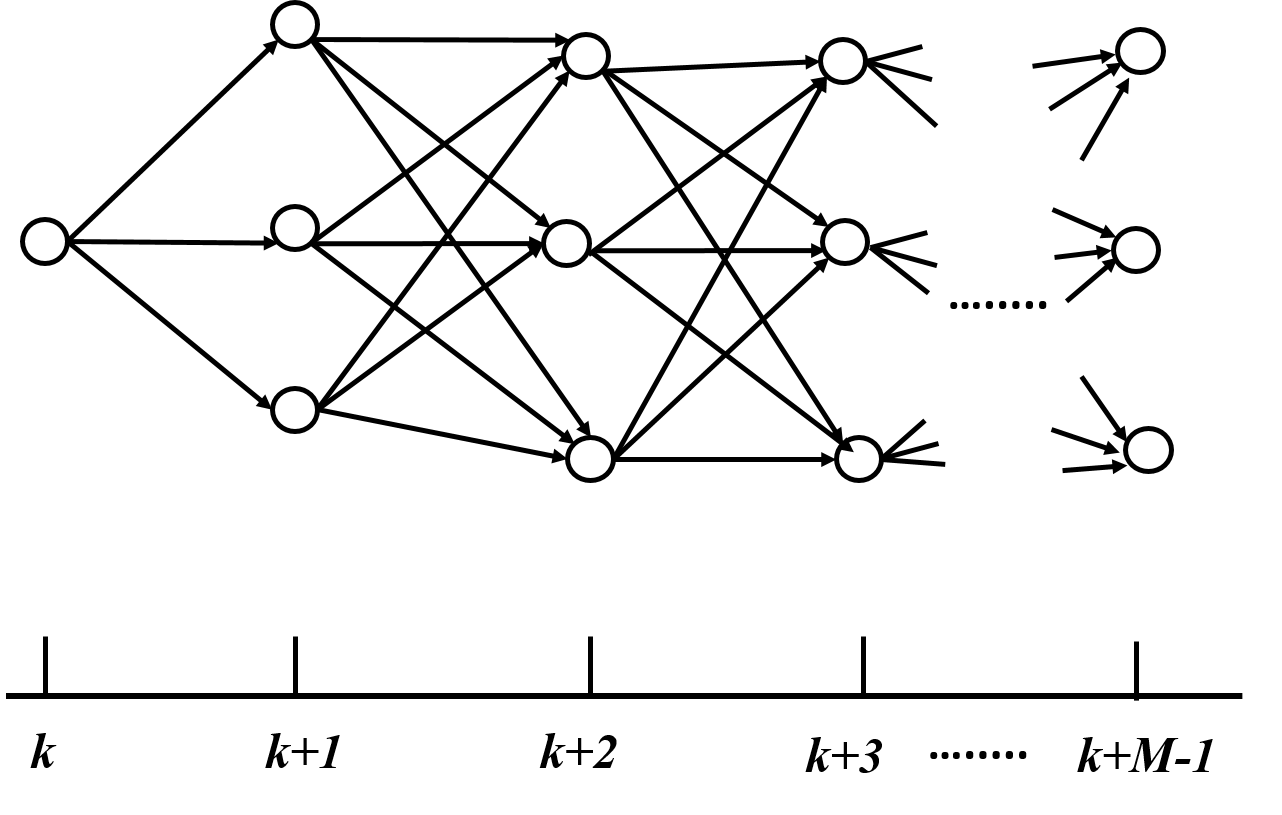}
\caption{Scenario tree with all possible scenarios}
\label{w_tree}
\end{figure}

For evaluating the expected cost in Eq. (\ref{cost}) over a finite horizon, scenario based stochastic model predictive control \cite{di2014stochastic} is used where a scenario is defined by a sequence of disturbance realizations $\{\omega_i(\cdot)\}$ over the given horizon. The number of such scenarios $n_{sc}$ would be exponential to the discrete time horizon $M$. The scenarios can be represented by a scenario tree (each path in the scenario tree represents a scenario) where the root node is the current $\omega_i(k)$ and the leaf nodes are the disturbance states reached at the end of discrete time horizon $M$. Fig. \ref{w_tree} shows the possible different scenarios when $|\mathcal{W}|$ is $3$. 

\begin{figure}[h]
%\graphicspath{{./fig/}}
\centering
\includegraphics[width=3in]{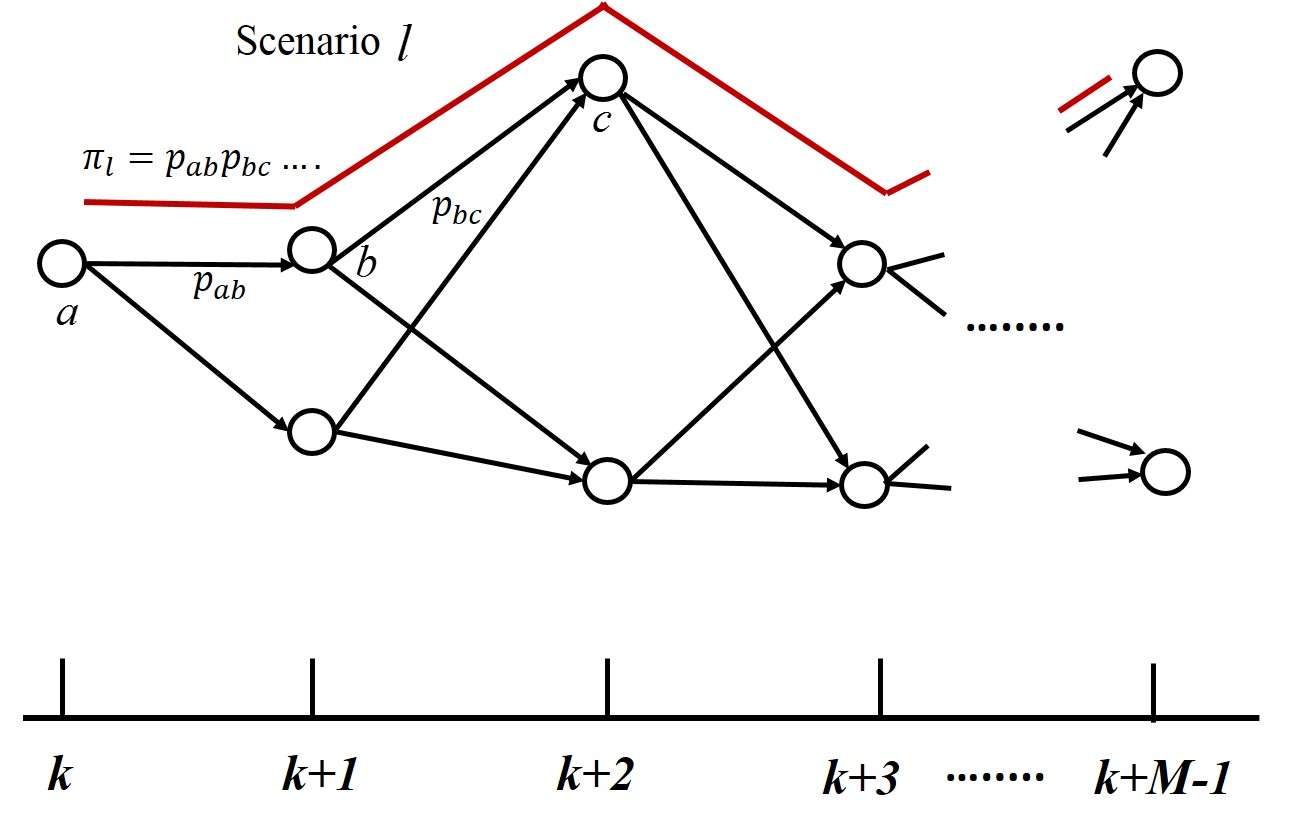}
\caption{Pruned scenario tree}
\label{n_tree}
\end{figure}

From the scenario tree, the cost function in Eq. (\ref{cost}) of the optimization problem in Eq. (\ref{cost1}) can now be represented as:
\begin{align}
 J_i(k) = \sum_{l=1}^{n_{sc}} \pi_l & \sum_{t=k}^{k+M-1}\bigg [c_1 \frac{\dot{Fuel}_i(t)\Delta t}{v_i(t) \Delta t} + c_2^i(t) R_{ij}(t)^2\notag\\
 &+ c_3 (v_i(t)  -v_{target}^i(k))^2 + {c}_4 u_i^f(t)^2 \bigg] \label{e_cost1}
\end{align}
Here, $n_{sc}$ is the total number of scenarios and $\pi_l$ is the probability of the occurrence of scenario (path) $l$ which is evaluated by the product of the probability values associated with all the edges in the path $l$. 
Since $n_{sc} = |\mathcal{W}|^M$, computation cost for solving the cost in Eq. (\ref{e_cost1}) would be huge and intractable.
To make the problem more computationally tractable, we use a sampling based method to generate a scenario tree where only the paths with higher probability of occurrence will be considered. The sampling based scenario tree generation method is explained in Algorithm \ref{algo_1} where $N_{max}<n_{sc}$ is the upper bound on the number of scenarios considered, $SC_l\{\cdot\} = \mathbb{R}^M$ contains the set of nodes in a scenario $l$, and $\mathcal{S}$ is the set of such $SC_l\{\cdot\}$ whose probability of occurrence is more than a predefined threshold $p_{th}$. Fig. \ref{n_tree} shows the pruned scenario tree along with the probability of occurrence of any path $l$.
\begin{algorithm}[h]
 \SetAlgoLined 
 \SetKwInOut{Input}{Input}\SetKwInOut{Output}{Output}
 \Input{Current state of $\omega_i(k)$: $a$ with probability 1, transition probability matrix $\mathcal{Q}_i$, threshold $p_{th}$, and $M = \frac{T}{\Delta t}$}
 %\Output{Set of nodes in every scenario $l$: $SC_l,\forall l$}
 \Output{Set of scenarios: $\mathcal{S}  = \{SC_l\{\cdot \} \} $ and their corresponding probability of occurrence $\pi_l$}
  Initialize $\mathcal{S} = \emptyset$ \;
  \For{$l = 1$ to $N_{max}$}{
 Initialize $SC_l\{1\} = a$, $\pi_l = 1$\;
   \For{$t_a =1$ to $M$}{
  	Choose $SC_l\{t_a + 1\} = b$ with probability $\mathcal{Q}_i(SC_l(t_a),b)$ \;
	$\pi_l = \pi_l \times \mathcal{Q}_i(SC_l(t_a),b)$ \;
 }
 \If{$ \pi_l\geq p_{th}$}{
 $\mathcal{S} = \{ \mathcal{S}, SC_l\{\cdot\} \}$ 
 } 
 }
 \caption{Sampling based scenario tree generation}
\label{algo_1}
\end{algorithm}

Now, the problem in Eq. (\ref{cost1}) can be expressed as
\begin{subequations}\label{ne_prob}
\begin{align} 
\min\limits_{u_i} \sum_{l=1}^{ |\mathcal{S}|} \pi_l & \sum_{t=k}^{k+M-1} \bigg[c_1 \frac{\dot{Fuel}_i(t)\Delta t}{v_i(t) \Delta t} + c_2^i(t) R_{ij}(t)^2\notag\\
 &+ c_3 (v_i(t)  -v_{target}^i(k))^2 + \bar{c}_4(t) u_i^f(t)^2\bigg] \label{e_cost2}\\
 \bar{c}_4(t) &=\frac{c_4}{(1+\alpha t)} \label{w_4_n} \\
 v_{min}&\leq v_i(t)\leq v_{max}  \label{cons1a}\\
u_{min}^i&\leq u_i^f(t)\leq u_{max}^i\label{cons2a}
\end{align}
\end{subequations}
%\begin{subequations}\label{ne_prob}
%\begin{align} 
%\min\limits_{u_i} \sum_{l=1}^{N_{max}} \pi_l & \sum_{t=k}^{M-1} [w_1 \frac{\dot{Fuel}_i(t)\Delta t}{v_i(t) \Delta t} + w_2^i(t) R_{ij}(t)^2\notag\\
% &+ w_3 (v_i(t)  -v_{target}^i(k))^2 + w_4^n(t) u_i^f(t)^2] \label{e_cost2}\\
% w_4(t) &=\frac{w_4}{(1+\alpha t)} \label{w_4_n} \\
% v_{min}&\leq v_i(k)\leq v_{max}, \forall k \label{cons1a}\\
%u_{min}^i&\leq u_i^f(k)\leq u_{max}^i,  \forall k \label{cons2a}
%\end{align}
%\end{subequations}
In the above problem, we have replaced $n_{sc}$ in Eq. (\ref{e_cost1}) with $|\mathcal{S}|< n_{sc}$ and we have changed the constant weight $c_4$ to a time varying weight $\bar{c}_4(t)$ given by Eq. (\ref{w_4_n}). By using such a time varying weight, we are discounting the future control efforts and penalizing the ones closer to the current time instant. This discounted cost is common in many different algorithms such as in reinforcement learning. It may be noted that by only summing over the number of scenarios $|\mathcal{S}|$ whose probability of occurrence is higher than $p_{th}$, we reduce the computational cost by a good amount. 
Thus the driver assistance system solves the problem in Eq. (\ref{ne_prob}) with the modified cost in a receding horizon fashion.
%and the constraints over the system dynamics and the ones presented in Eq. (\ref{cost1}) and Eq. (\ref{cost_old1}). In this paper, we consider each vehicle $j$ shares its future position and velocity trajectory, obtained from solving its finite horizon problem, to the vehicle $i$ just behind. 

%%%%%%%%%%%%%%%%%%%%%%%%%%%%%%%%%%%%%%%%%%%%%%%%%%%%%%%%%%%%%%%%%%%%%%%%%%%%%%
%\FloatBarrier
\section{SIMULATION RESULTS}\label{simulation}
In this section, we present the simulation results of the methodology explained in the previous Section \ref{approach}. We consider a simulation scenario consisting of a single lane road with traffic lights at every $500$m. Each vehicle implements a scenario based stochastic model predictive controller by solving problem in Eq. (\ref{ne_prob}) so that: i) each vehicle improves its fuel efficiency, ii) avoids rear end collision with the preceding vehicle, iii) reduces red light idling, and iv) Addresses error introduced by human driver.

We have considered $|\mathcal{W}| = 5$, so that the transition probability matrix is $\mathcal{Q}_i\in \mathbb{R}^{5\times 5}$. The transition probability matrix captures the driver behavior. For example, if the error is positive, an aggressive driver might overcompensate and make the error negative rather than zero. Ideally, $\mathcal{W}$ and $\mathcal{Q}_i$ needs to be generated and updated analyzing a driver's ability in tracking commands from the driver assistance system. Since experiments on real vehicles are beyond the scope of this paper, we assume the data already exists to evaluate our proposed method. We consider two simulation scenarios in this paper: (i) $\mathcal{W}$ ranges from $-0.3$ to $0.3$, and (ii) $\mathcal{W}$ ranges from $-0.5$ to $0.5$. 

First, we compare our proposed method with the ideal case where the driver induced error is zero and then with a passive controller (baseline method) where the driver assistance system does not address the driver induced error (considers the driver follows its instructions exactly). In both ideal and the baseline cases, the controller solves the problem in Eq. \ref{cost1}, without the expectation, using model predictive control. For the ideal case, $\omega_i(k) = 0, \forall k$ while for the baseline case, $\omega_i(k)$ follows the Markov chain but the method assumes it to be zero.

We then compare our proposed method with two optimal control methods for stochastic systems: certainty equivalence control \cite{stengel1986stochastic} and frozen time model predictive control \cite{di2014stochastic,bernardini2012stabilizing}. In certainty equivalence optimal control problem, the future random error realizations are assumed to be known and often replaced by the expected error. In frozen time model predictive control, the future error realizations (over the horizon) are considered to be constant and equal to the initial error. 
A number of papers (including our previous works) have already shown how avoiding red light idling, using V2V and V2I information, and developing controls in a model predictive framework improves vehicle fuel efficiency. Thus we are not comparing our method with the traditional car following models.

For both the scenarios, we run the simulation for $400$ seconds. For simulation, most of the parameters are taken from \cite{kamal2010ecological} and \cite{kamal2013model}. The vehicles are considered to be identical with $A_v^i = 2.5$ $m^2$, $M_h^i = 1200$ kg and $u_{max}^i =2$ $m/s^2, \forall i$. The other parameters are given by: $C_D = 0.32$, $\rho_a = 1.184$ $kg/m^3$, and $\theta = 0$ degree. The parameters used in fuel consumption model are: $b_0 = 0.1569$, $b_1 = 2.450\times 10^{-2}$, $b_2 = -7.415\times 10^{-4}$, $b_3=5.975\times10^{-5}$, $c_0 = 0.07224$, $c_1 = 9.681\times 10^{-2}$, $c_2 = 1.075\times 10^{-3}$ and $F_d^i = 0.1, \forall i$ \cite{kamal2010ecological}. To emulate urban road conditions, the red $t_r$ and green light $t_g$  intervals are sampled from a uniform distribution with range $37$ to $43$ and $12$ to $17$ seconds respectively for each cycle of every traffic signal. The maximum ($v_{max}$) and minimum ($v_{min}$) allowable velocities are considered to be $20$ m/s  and $0$ m/s respectively. For simulation purposes, time horizon is chosen as $T = 5$ secs with time step $\Delta t = 0.5 $ secs (thus $M = 10$) and $n=3$ vehicles are considered.

\textbf{Scenario 1}: Here, $\omega_i(\cdot)$ is considered to take discrete values $[-0.3\quad -0.15 \quad 0 \quad 0.15 \quad 0.3], \forall i$. Although $\omega_i(\cdot)$ is discretized into $5$ discrete values for solving the stochastic model predictive control problem, the actual driver error would take any value in $-0.3 \leq \omega_i(\cdot)\leq 0.3$.

The state transition matrix considered in this paper is given by Eq. (\ref{tau_i}).
\begin{align}\label{tau_i}
\mathcal{Q}_i = \left [ \begin{array}{ccccc}
0.10 & 0.30 & 0.45 & 0.10 & 0.05 \\
0.05 & 0.25 & 0.45 & 0.20 & 0.05 \\
0.01 & 0.10 & 0.78  &  0.10  &  0.01 \\
0.05 & 0.15 & 0.45  & 0.25 & 0.10 \\
 0.05 & 0.15 & 0.45 & 0.30 & 0.05
 \end{array} \right] 
\end{align}

\begin{figure}[h!]
%\graphicspath{{./fig/}}
\centering
\subfloat[Vehicle 1]{\label{fig2a}\includegraphics[width=2.4in]{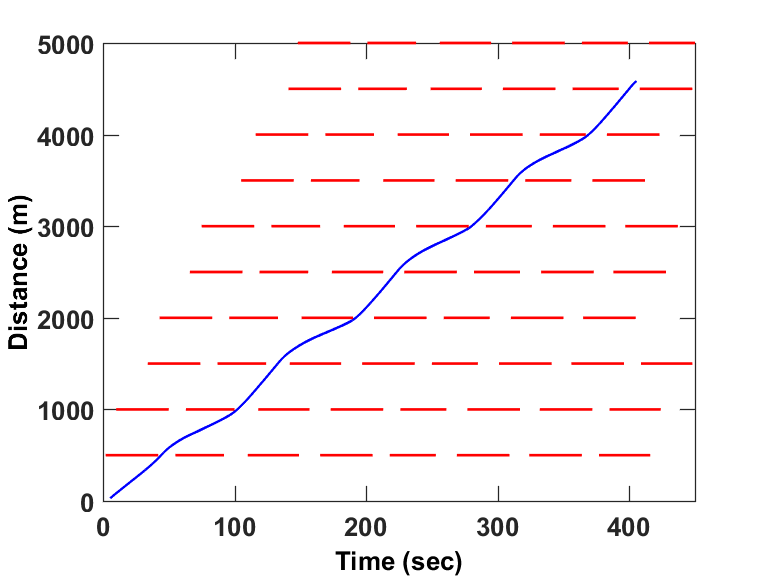}} \quad
\subfloat[Vehicle 2]{\label{fig2b}\includegraphics[width=2.4in]{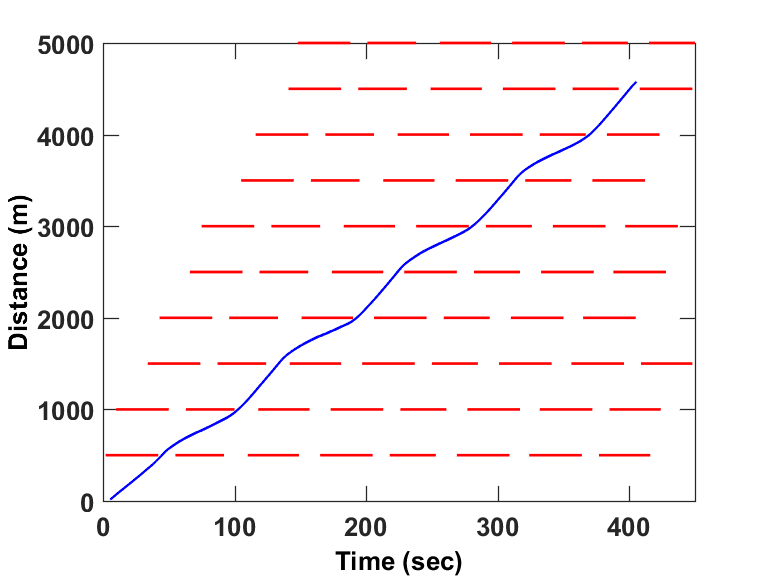}} \quad
\subfloat[Vehicle 3]{\label{fig2c}\includegraphics[width=2.4in]{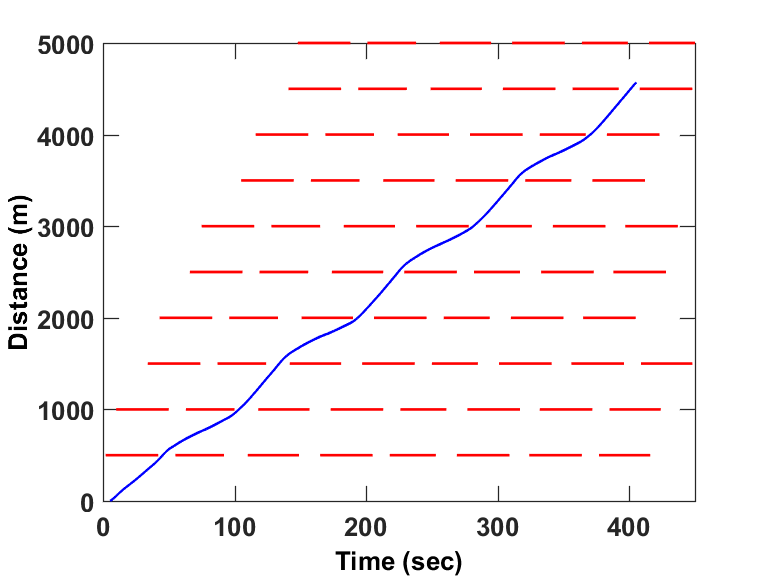}}
\caption{Vehicle trajectories of the 3 vehicles following our proposed method}
\label{fig2}
\end{figure}

\begin{figure}[h]
%\graphicspath{{./fig/}}
\centering
\subfloat[Vehicle 1]{\label{fig4a}\includegraphics[width=2.5in]{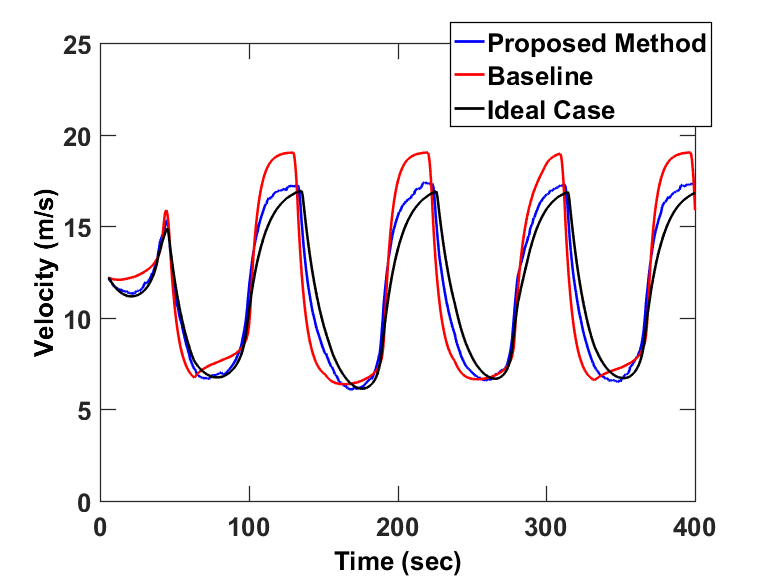}} \quad
\subfloat[Vehicle 2]{\label{fig4b}\includegraphics[width=2.5in]{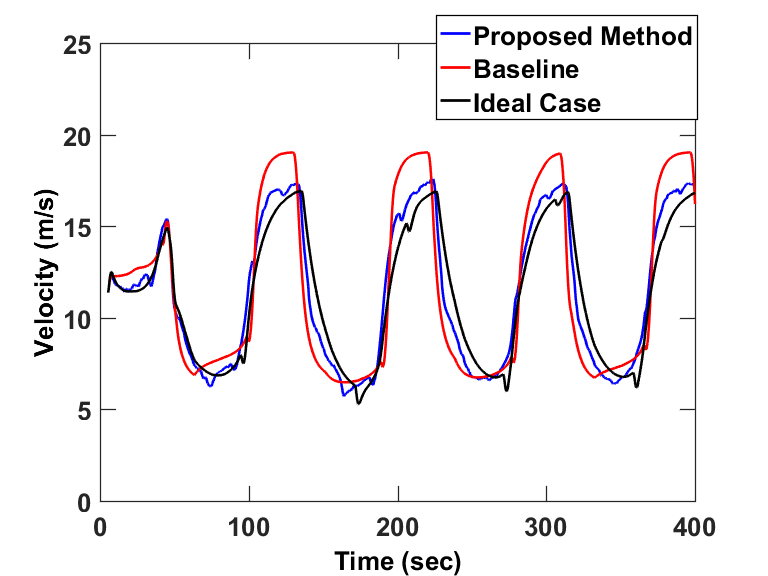}} \quad
\subfloat[Vehicle 3]{\label{fig4c}\includegraphics[width=2.5in]{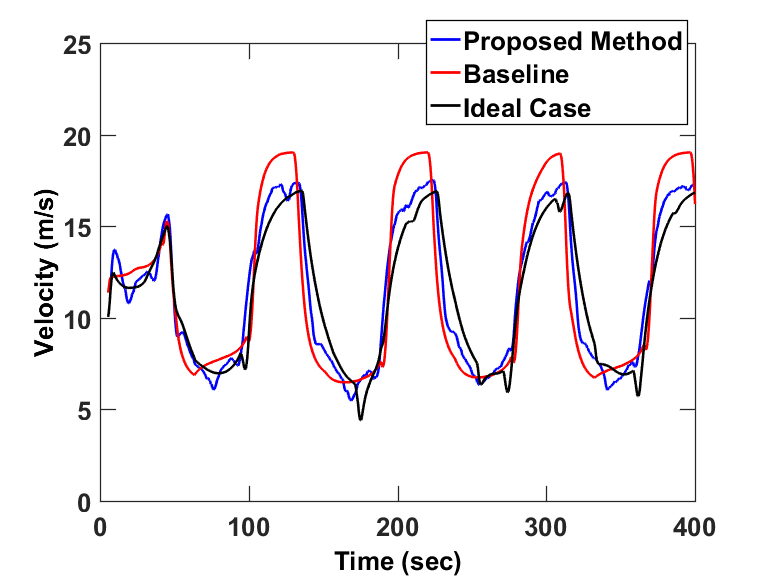}}
\caption{Comparison of velocity profiles between ideal case, our proposed method, and the baseline case (passive controller) for Scenario 1}
\label{fig4}
\end{figure}

Fig. \ref{fig2} shows the trajectories of the three vehicles using our proposed method and how they all avoid red light idling. The red bars in Fig. \ref{fig2} are the red light intervals and the blank spaces are the green light intervals so that the vehicles need to pass a traffic light through the blank spaces. The vehicle trajectories of all the $3$ vehicles look similar since they end up following each other in a fuel economic manner and no collision takes place. Fig. \ref{fig4} shows the comparison between the velocity profiles for the 3 vehicles in the ideal case, when using our proposed method, and for the passive controller. It can be seen from Fig. \ref{fig4} that the velocity profiles generated following our proposed method are much closer to the ones in the ideal case. This is reflected in the fuel economy of the vehicles, as it can be seen from Tab. \ref{tab1} that the fuel economy of the vehicles following our proposed method are much closer to the fuel economy in the ideal case. 
%Tab. \ref{tab11} shows the standard deviation of the control error, which is the difference between the control output for the ideal case and with driver imperfection given by $u^{ideal}_i - u^f_i,\forall i$, for all the vehicles when using baseline and our proposed method. It can be seen from the Tab. \ref{tab11}, that the error standard deviation is significantly reduced when using our proposed approach.

%Since the system is stochastic in nature, different simulation runs with the same initial conditions would give different results. For the Scenario 1, we ran the simulation for $10$ different runs with the same initial conditions and the fuel efficiency, measured as miles per gallon, for each run is shown in Fig. \ref{fig4_1}. For our proposed method the standard deviation of the fuel efficiency for the $10$ runs are: for vehicle 1 it is  is $0.16$ mpg; for vehicle 2 it is $0.23$ mpg; and for vehicle 3 it is $0.21$ mpg. It can also be seen from Fig. \ref{fig4_1a}, Fig. \ref{fig4_1b} and Fig. \ref{fig4_1c} that the fuel efficiency for all the different runs are better when using our proposed method as opposed to the baseline method for vehicles 1, vehicle 2 and vehicle 3 respectively.

%\begin{figure}[h]
%%\graphicspath{{./fig/}}
%\centering
%\subfloat[Vehicle 1]{\label{fig3a}\includegraphics[width=2.4in]{prob_sc1_acc1.eps}} \quad
%\subfloat[Vehicle 2]{\label{fig3b}\includegraphics[width=2.4in]{prob_sc1_acc2.eps}} \quad
%\subfloat[Vehicle 3]{\label{fig3c}\includegraphics[width=2.4in]{prob_sc1_acc3.eps}}
%\caption{Vehicle accelerations of the 3 vehicles following our proposed method for Scenario 1}
%\label{fig3}
%\end{figure}

\begin{table}[h]
\caption{Individual vehicle fuel economy (mpg) after $400$ seconds for Scenario 1}
\label{tab1}
\centering
%\scriptsize
\begin{tabular}{|c||c||c||c|}
\hline
Vehicle No. & Ideal Case  & Passive Controller & Proposed Method \\ 
\hline
1 & 41.52 & 37.51   & 40.65  \\
2 &  40.58 &  37.56 & 39.65\\
3 & 40.08 &  36.78 & 39.32 \\
\hline 
\end{tabular}
\end{table}

%\begin{table}[h]
%\caption{Standard Deviation of Control Solution Difference Scenario 1}
%\label{tab11}
%\centering
%%\scriptsize
%\begin{tabular}{|c||c||c|}
%\hline
%Vehicle No. &  Baseline & Proposed Method \\ 
%\hline
%1 & 0.38 & 0.17  \\
%2 & 0.46 & 0.22\\
%3 &  0.49 & 0.27 \\
%\hline 
%\end{tabular}
%\end{table}
\begin{figure}[h]
%\graphicspath{{./fig/}}
\centering
\subfloat[Vehicle 1]{\label{fig4_1a}\includegraphics[width=2.7in]{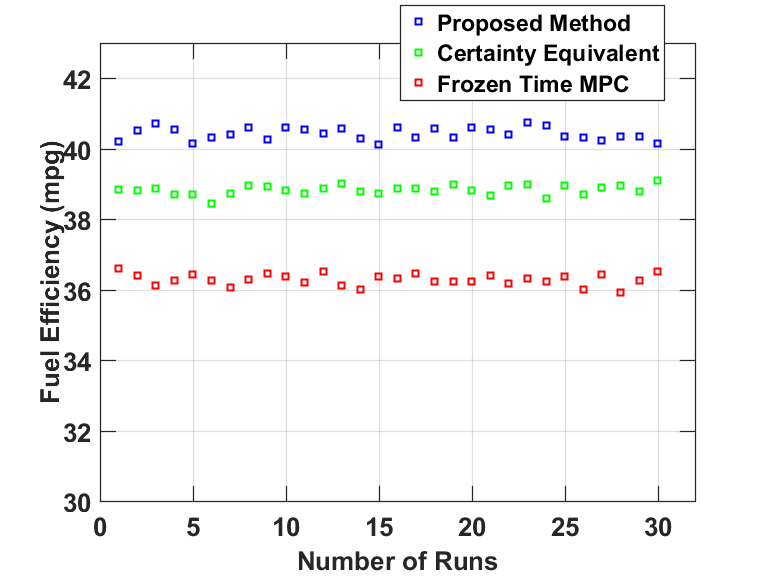}} \quad
\subfloat[Vehicle 2]{\label{fig4_1b}\includegraphics[width=2.7in]{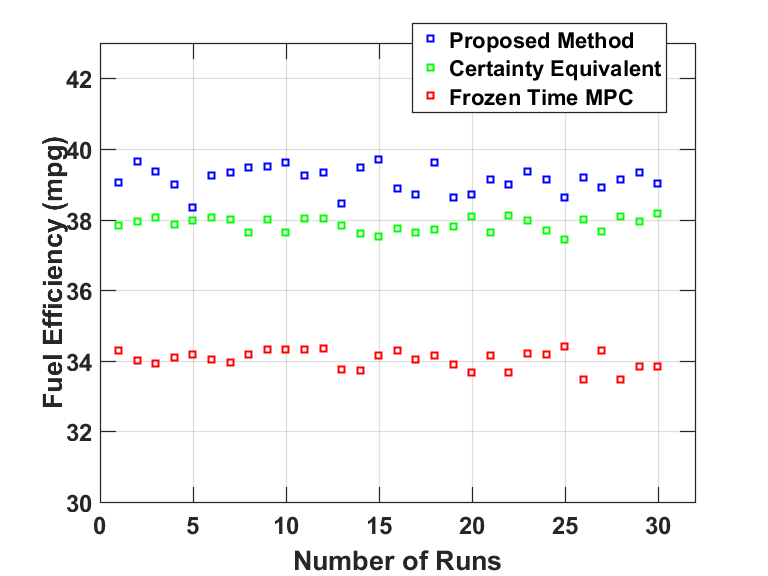}} \quad
\subfloat[Vehicle 3]{\label{fig4_1c}\includegraphics[width=2.7in]{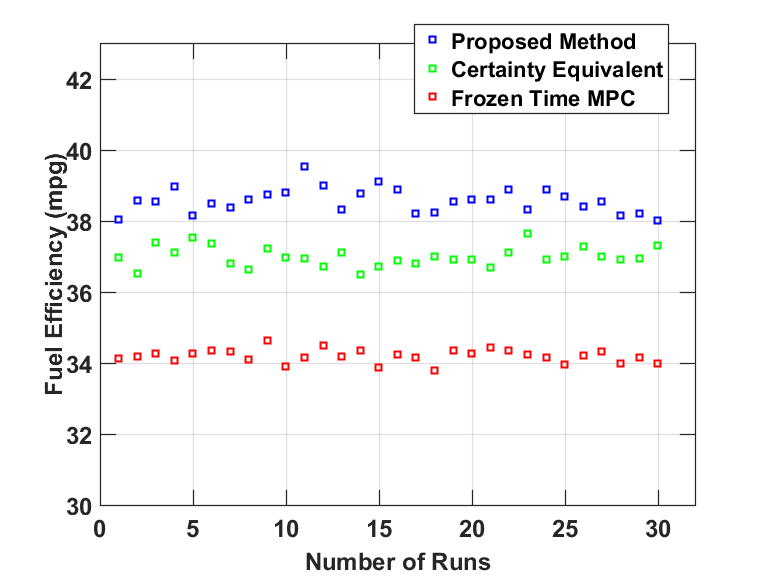}}
\caption{Comparison of miles per gallon for 30 cases when using our proposed method, certainty equivalence control, and frozen time model predictive control for Scenario 1}
\label{fig4_1}
\end{figure}

We next compare our proposed method with two different optimal control methods for uncertain systems. Since the system is stochastic in nature, different simulation runs with the same initial conditions would give different results. That is why we ran $30$ different simulation runs to compare our proposed method with certainty equivalence optimal control and frozen time model predictive control. The results are shown in Fig. \ref{fig4_1}. It can be seen from the Fig. \ref{fig4_1} that certainty equivalence optimal control performs better that frozen time model predictive control while our proposed method out performs the other two methods.
%\begin{table}[h]
%\caption{Individual vehicle fuel economy (mpg) after $400$ seconds for Scenario 1}
%\label{tab_30}
%\centering
%%\scriptsize
%\begin{tabular}{|c||c||c||c|}
%\hline
%Vehicle No. & Certainty Equivalence   & Frozen Time MPC & Proposed Method \\ 
%\hline
%1 & 38.82 & 36.23   & 40.42  \\
%2 &  37.85 &  34.03 & 39.45\\
%3 & 36.98 &  34.19 & 39.10 \\
%\hline 
%\end{tabular}
%\end{table}

\begin{figure}[h]
%\graphicspath{{./fig/}}
\centering
\subfloat[Vehicle 1]{\label{fig5a}\includegraphics[width=2.5in]{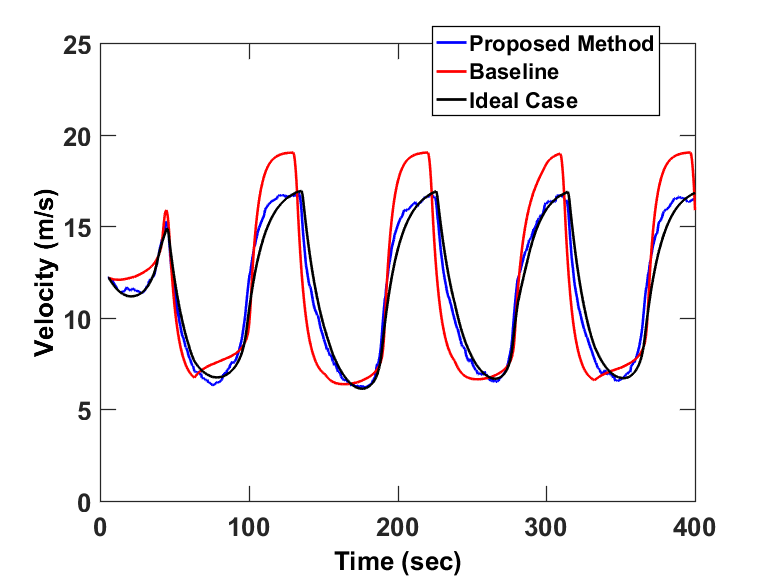}} \quad
\subfloat[Vehicle 2]{\label{fig5b}\includegraphics[width=2.5in]{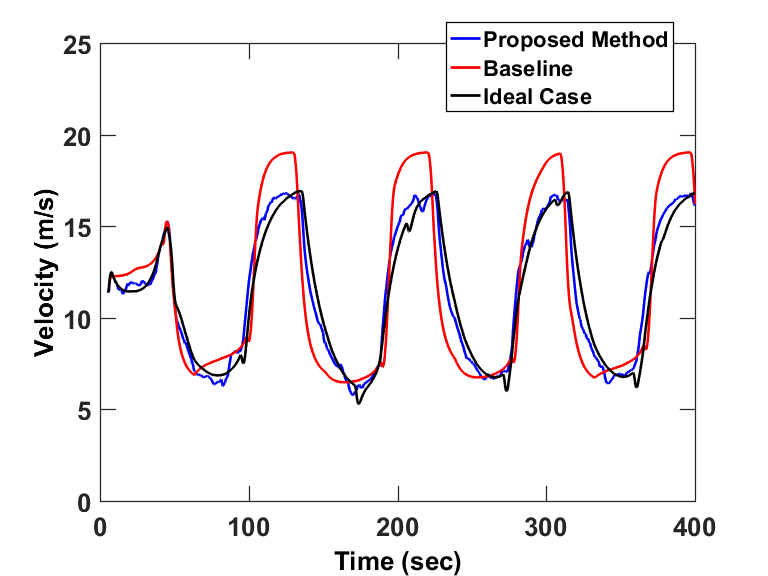}} \quad
\subfloat[Vehicle 3]{\label{fig5c}\includegraphics[width=2.5in]{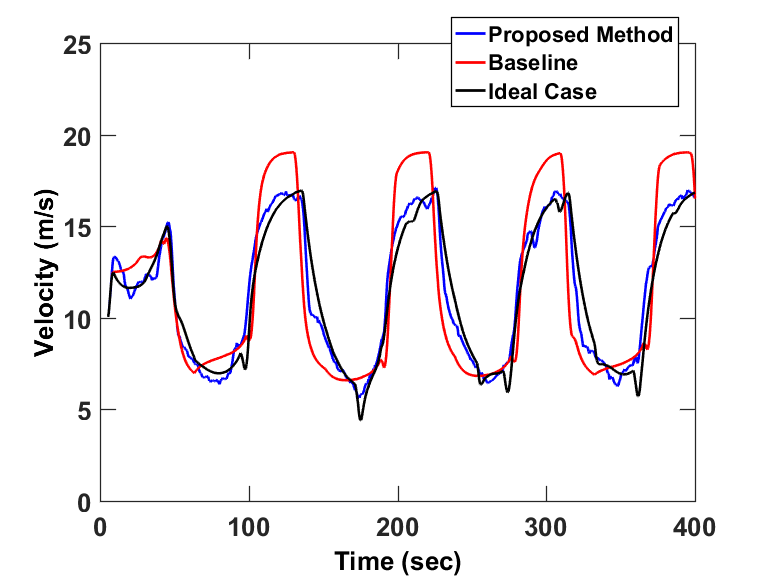}}
\caption{Comparison of velocity profiles between ideal case, our proposed method, and the baseline case (passive controller) for Scenario 2}
\label{fig5}
\end{figure}

\textbf{Scenario 2}: Here, $\omega_i$ is considered to take discrete values $[-0.5 \quad -0.2 \quad 0.0 \quad 0.2 \quad 0.5], \forall i$, while $\mathcal{Q}_i$ is given by Eq. (\ref{tau_i}). Fig. \ref{fig5} shows the comparison between the velocity profiles for the 3 vehicles in the ideal case, when using our proposed method, and for the passive controller (baseline case). Similar to the previous case, it can be seen from Fig. \ref{fig5} that the velocity profiles generated following our proposed method are much closer to the ones in the ideal case. This is reflected in the fuel economy of the vehicles, as it can be seen from Tab. \ref{tab2} that the fuel economy of the vehicles following our proposed method are much closer to the fuel economy in the ideal case. 
%Similar to the previous scenario, Tab. \ref{tab22} shows significant reduction of control error standard deviation from our proposed method.

\begin{table}[h]
\caption{Individual vehicle fuel economy (mpg) after $400$ seconds for Scenario 2}
\label{tab2}
\centering
%\scriptsize
\begin{tabular}{|c||c||c||c|}
\hline
Vehicle No. & Ideal Case & Passive Controller & Proposed Method \\ 
\hline
1 & 41.52 & 36.90  & 40.28  \\
2 &  40.58 &   36.35 &  39.38 \\
3 & 40.08 &   36.17 &  39.12\\

\hline
\end{tabular}
\end{table}

%\begin{table}[h]
%\caption{Standard Deviation of Control Solution Difference Scenario 2}
%\label{tab22}
%\centering
%%\scriptsize
%\begin{tabular}{|c||c||c|}
%\hline
%Vehicle No. &  Baseline & Proposed Method \\ 
%\hline
%1 & 0.39 & 0.19  \\
%2 & 0.48 & 0.25\\
%3 &  0.51 & 0.31 \\
%\hline 
%\end{tabular}
%\end{table}

\section{CONCLUSIONS} \label{conclusion}
In this paper, we present a driver-in the loop fuel efficient control strategy for a group of CVs in urban road conditions. The control strategy works as a driver assistance system where the driver behavior and its capability is considered in controller development. Anticipation of driver behavior improves the system performance as shown by the simulation results. The strategy works in a decentralized manner as every vehicle develops its own control strategy using neighborhood V2V and V2I information. Utilization of SPAT information via V2I communication improves system mobility and in the process reduces green house emissions. Some of the future research directions include improvement of computational requirement in solving the scenario based stochastic model predictive control problem and consideration of network delay and communication loss in the development of control strategy for the CVs. 
%Future research direction also includes development of fuel efficient control strategies for multi-lane roads with intersections.


\clearpage
%\bibliographystyle{ieeetr}

%\FloatBarrier
%\bibliography{ref_j4}
\bibliography{}

\begin{thebibliography}{10}

\bibitem{2010cong}
F.~T.~A. U.S.Department~of Transportation, Federal Highway~Administration,
  ``2010 status of the nation�s highways, bridges, and transit: Conditions
  and performances,'' tech. rep., Report to Congress, 2010.

\bibitem{najm2010frequency}
W.~G. Najm, J.~Koopmann, J.~D. Smith, and J.~Brewer, ``Frequency of target
  crashes for intellidrive safety systems,'' tech. rep., U.S. Department of
  Transportation � National Highway Traffic Safety Administration, 2010.

\bibitem{van2004driving}
J.~Van~Mierlo, G.~Maggetto, E.~Van~de Burgwal, and R.~Gense, ``Driving style
  and traffic measures-influence on vehicle emissions and fuel consumption,''
  {\em Proceedings of the Institution of Mechanical Engineers, Part D: Journal
  of Automobile Engineering}, vol.~218, no.~1, pp.~43--50, 2004.

\bibitem{gilbert1976vehicle}
E.~G. Gilbert, ``Vehicle cruise: improved fuel economy by periodic control,''
  {\em Automatica}, vol.~12, no.~2, pp.~159--166, 1976.

\bibitem{chang2005vehicle}
D.~J. Chang and E.~K. Morlok, ``Vehicle speed profiles to minimize work and
  fuel consumption,'' {\em Journal of Transportation Engineering}, vol.~131,
  no.~3, pp.~173--182, 2005.

\bibitem{hooker1988optimal}
J.~Hooker, ``Optimal driving for single-vehicle fuel economy,'' {\em
  Transportation Research Part A: General}, vol.~22, no.~3, pp.~183--201, 1988.

\bibitem{kirschbaum2002determination}
F.~Kirschbaum, M.~Back, and M.~Hart, ``Determination of the fuel-optimal
  trajectory for a vehicle along a known route,'' in {\em Proceedings of the
  15th IFAC World Congress}, vol.~15, pp.~1505--1505, 2002.

\bibitem{hellstrom2010design}
E.~Hellstr{\"o}m, J.~{\AA}slund, and L.~Nielsen, ``Design of an efficient
  algorithm for fuel-optimal look-ahead control,'' {\em Control Engineering
  Practice}, vol.~18, no.~11, pp.~1318--1327, 2010.

\bibitem{kamal2010ecological}
M.~Kamal, M.~Mukai, J.~Murata, and T.~Kawabe, ``Ecological driver assistance
  system using model-based anticipation of vehicle--road--traffic
  information,'' {\em IET Intelligent Transport Systems}, vol.~4, no.~4,
  pp.~244--251, 2010.

\bibitem{kamal2013model}
M.~S. Kamal, M.~Mukai, J.~Murata, and T.~Kawabe, ``Model predictive control of
  vehicles on urban roads for improved fuel economy,'' {\em IEEE Transactions
  on Control Systems Technology}, vol.~21, no.~3, pp.~831--841, 2013.

\bibitem{asadi2011predictive}
B.~Asadi and A.~Vahidi, ``Predictive cruise control: Utilizing upcoming traffic
  signal information for improving fuel economy and reducing trip time,'' {\em
  IEEE Transactions on Control Systems Technology}, vol.~19, no.~3,
  pp.~707--714, 2011.

\bibitem{mahler2012reducing}
G.~Mahler and A.~Vahidi, ``Red light avoidance through probabilistic traffic
  signal timing prediction,'' {\em IEEE Transactions on Intelligent
  Transportation Systems}, vol.~15, pp.~2516-- 2523, 2014.

\bibitem{de2013eco}
G.~De~Nunzio, C.~Canudas~de Wit, P.~Moulin, and D.~Di~Domenico, ``Eco-driving
  in urban traffic networks using traffic signal information,'' in {\em 2013
  IEEE 52nd Annual Conference on Decision and Control (CDC)}, pp.~892--898,
  IEEE, 2013.

\bibitem{mandava2009arterial}
S.~Mandava, K.~Boriboonsomsin, and M.~Barth, ``Arterial velocity planning based
  on traffic signal information under light traffic conditions,'' in {\em 12th
  International IEEE Conference on Intelligent Transportation Systems, 2009.
  ITSC'09}, pp.~160--165, IEEE, 2009.

\bibitem{rakha2011eco}
H.~Rakha and R.~K. Kamalanathsharma, ``Eco-driving at signalized intersections
  using v2i communication,'' in {\em 14th International IEEE Conference on
  Intelligent Transportation Systems (ITSC)}, pp.~341--346, IEEE, 2011.

\bibitem{kamalanathsharma2013multi}
R.~K. Kamalanathsharma and H.~Rakha, ``Multi-stage dynamic programming
  algorithm for eco-speed control at traffic signalized intersections,'' in
  {\em 16th International IEEE Conference on Intelligent Transportation
  Systems-(ITSC)}, pp.~2094--2099, IEEE, 2013.

\bibitem{ozatay2013analytical}
E.~Ozatay, U.~Ozguner, D.~Filev, and J.~Michelini, ``Analytical and numerical
  solutions for energy minimization of road vehicles with the existence of
  multiple traffic lights,'' in {\em Decision and Control (CDC), 2013 IEEE 52nd
  Annual Conference on}, pp.~7137--7142, IEEE, 2013.

\bibitem{xia2012field}
H.~Xia, K.~Boriboonsomsin, F.~Schweizer, A.~Winckler, K.~Zhou, W.-B. Zhang, and
  M.~Barth, ``Field operational testing of eco-approach technology at a
  fixed-time signalized intersection,'' in {\em 15th International IEEE
  Conference on Intelligent Transportation Systems (ITSC)}, pp.~188--193, IEEE,
  2012.

\bibitem{ozataybayesian}
E.~Ozatay, U.~Ozguner, D.~Filev, and J.~Michelini, ``Bayesian traffic-light
  parameter tracking based on semihidden markov models,''

\bibitem{bengler2014three}
K.~Bengler, K.~Dietmayer, B.~Farber, M.~Maurer, C.~Stiller, and H.~Winner,
  ``Three decades of driver assistance systems: Review and future
  perspectives,'' {\em Intelligent Transportation Systems Magazine, IEEE},
  vol.~6, no.~4, pp.~6--22, 2014.

\bibitem{hummel2011advanced}
T.~Hummel, M.~K{\"u}hn, J.~Bende, and A.~Lang, ``Advanced driver assistance
  systems: an investigation of their potential safety benefits based on an
  analysis of insurance claims in germany,'' {\em German Insurance Association
  Insurers Accident Research, Research Report FS}, vol.~3, 2011.

\bibitem{shladover2012literature}
S.~Shladover, ``Literature review on recent international activity in
  cooperative vehicle--highway automation systems,'' tech. rep., 2012.

\bibitem{piao2008advanced}
J.~Piao and M.~McDonald, ``Advanced driver assistance systems from autonomous
  to cooperative approach,'' {\em Transport Reviews}, vol.~28, no.~5,
  pp.~659--684, 2008.

\bibitem{wu2011fuel}
C.~Wu, G.~Zhao, and B.~Ou, ``A fuel economy optimization system with
  applications in vehicles with human drivers and autonomous vehicles,'' {\em
  Transportation Research Part D: Transport and Environment}, vol.~16, no.~7,
  pp.~515--524, 2011.

\bibitem{vagg2013development}
C.~Vagg, C.~J. Brace, D.~Hari, S.~Akehurst, J.~Poxon, and L.~Ash, ``Development
  and field trial of a driver assistance system to encourage eco-driving in
  light commercial vehicle fleets,'' {\em Intelligent Transportation Systems,
  IEEE Transactions on}, vol.~14, no.~2, pp.~796--805, 2013.

\bibitem{gilman2015personalised}
E.~Gilman, A.~Keskinarkaus, S.~Tamminen, S.~Pirttikangas, J.~R{\"o}ning, and
  J.~Riekki, ``Personalised assistance for fuel-efficient driving,'' {\em
  Transportation Research Part C: Emerging Technologies}, vol.~58,
  pp.~681--705, 2015.

\bibitem{guan2012fuel}
T.~Guan and C.~W. Frey, ``Fuel efficiency driver assistance system for
  manufacturer independent solutions,'' in {\em Intelligent Transportation
  Systems (ITSC), 2012 15th International IEEE Conference on}, pp.~212--217,
  IEEE, 2012.

\bibitem{di2014stochastic}
S.~Di~Cairano, D.~Bernardini, A.~Bemporad, and I.~V. Kolmanovsky, ``Stochastic
  mpc with learning for driver-predictive vehicle control and its application
  to hev energy management,'' {\em Control Systems Technology, IEEE
  Transactions on}, vol.~22, no.~3, pp.~1018--1031, 2014.

\bibitem{gray2013stochastic}
A.~Gray, Y.~Gao, T.~Lin, J.~K. Hedrick, and F.~Borrelli, ``Stochastic
  predictive control for semi-autonomous vehicles with an uncertain driver
  model,'' in {\em Intelligent Transportation Systems-(ITSC), 2013 16th
  International IEEE Conference on}, pp.~2329--2334, IEEE, 2013.

\bibitem{shia2014semiautonomous}
V.~A. Shia, Y.~Gao, R.~Vasudevan, K.~D. Campbell, T.~Lin, F.~Borrelli, and
  R.~Bajcsy, ``Semiautonomous vehicular control using driver modeling,'' {\em
  IEEE Transactions on Intelligent Transportation Systems}, vol.~15, no.~6,
  pp.~2696--2709, 2014.

\bibitem{gray2012semi}
A.~Gray, M.~Ali, Y.~Gao, J.~Hedrick, and F.~Borrelli, ``Semi-autonomous vehicle
  control for road departure and obstacle avoidance,'' {\em IFAC control of
  transportation systems}, pp.~1--6, 2012.

\bibitem{bernardini2009scenario}
D.~Bernardini and A.~Bemporad, ``Scenario-based model predictive control of
  stochastic constrained linear systems,'' in {\em Decision and Control, 2009
  held jointly with the 2009 28th Chinese Control Conference. CDC/CCC 2009.
  Proceedings of the 48th IEEE Conference on}, pp.~6333--6338, IEEE, 2009.

\bibitem{grosso2013assessment}
J.~M. Grosso, J.~M. Maestre, C.~Ocampo-Martinez, and V.~Puig, ``On the
  assessment of tree-based and chance-constrained predictive control approaches
  applied to drinking water networks,'' IFAC, 2013.

\bibitem{leidereiter2014quadrature}
C.~Leidereiter, A.~Potschka, and H.~G. Bock, ``Quadrature-based scenario tree
  generation for nonlinear model predictive control,'' in {\em Proc. of IFAC
  World Congress Cape Town}, pp.~11087--11092, 2014.

\bibitem{heitsch2009scenario}
H.~Heitsch and W.~R{\"o}misch, ``Scenario tree modeling for multistage
  stochastic programs,'' {\em Mathematical Programming}, vol.~118, no.~2,
  pp.~371--406, 2009.

\bibitem{bernardini2012stabilizing}
D.~Bernardini and A.~Bemporad, ``Stabilizing model predictive control of
  stochastic constrained linear systems,'' {\em IEEE Transactions on Automatic
  Control}, vol.~57, no.~6, pp.~1468--1480, 2012.

\bibitem{HomChaudhuri2015}
B.~HomChaudhuri, A.~Vahidi, and P.~Pisu, ``A fuel economic model predictive
  control strategy for a group of connected vehicles in urban roads,'' in {\em
  American Control Conference}, pp.~2741--2746, 2015.

\bibitem{homchaudhuri2016hierarchical}
B.~HomChaudhuri, R.~Lin, and P.~Pisu, ``Hierarchical control strategies for
  energy management of connected hybrid electric vehicles in urban roads,''
  {\em Transportation Research Part C: Emerging Technologies}, vol.~62,
  pp.~70--86, 2016.

\bibitem{lin2015fuel}
R.~Lin, B.~HomChaudhuri, and P.~Pisu, ``Fuel efficient control strategies for
  connected hybrid electric vehicles in urban roads,'' in {\em ASME 2015
  Dynamic Systems and Control Conference}, pp.~1--7, American Society of
  Mechanical Engineers, 2015.

\bibitem{HomChaudhuri2016}
B.~HomChaudhuri, A.~Vahidi, and P.~Pisu, ``Fast model predictive control based
  fuel efficient control strategy for a group of connected vehicles in urban
  road conditions,'' {\em to appear in IEEE Transactions on Control Systems
  Technology}, 2016.

\bibitem{stengel1986stochastic}
R.~F. Stengel, {\em Stochastic optimal control}.
\newblock John Wiley and Sons New York, New York, 1986.

\bibitem{pentland1999modeling}
A.~Pentland and A.~Liu, ``Modeling and prediction of human behavior,'' {\em
  Neural computation}, vol.~11, no.~1, pp.~229--242, 1999.

\bibitem{takano2008recognition}
W.~Takano, A.~Matsushita, K.~Iwao, and Y.~Nakamura, ``Recognition of human
  driving behaviors based on stochastic symbolization of time series signal,''
  in {\em Intelligent Robots and Systems, 2008. IROS 2008. IEEE/RSJ
  International Conference on}, pp.~167--172, IEEE, 2008.

\bibitem{wang2009hmm}
Z.~Wang, A.~Peer, and M.~Buss, ``An hmm approach to realistic haptic
  human-robot interaction,'' in {\em EuroHaptics conference, 2009 and Symposium
  on Haptic Interfaces for Virtual Environment and Teleoperator Systems. World
  Haptics 2009. Third Joint}, pp.~374--379, IEEE, 2009.

\bibitem{lam2014pomdp}
C.-P. Lam and S.~S. Sastry, ``A pomdp framework for human-in-the-loop system,''
  in {\em Decision and Control (CDC), 2014 IEEE 53rd Annual Conference on},
  pp.~6031--6036, IEEE, 2014.

\bibitem{gray2002markov}
R.~Gray, ``�markov at the bat�: A model of cognitive processing in baseball
  batters,'' {\em Psychological Science}, vol.~13, no.~6, pp.~542--547, 2002.

\bibitem{sycara2015abstraction}
K.~Sycara, C.~Lebiere, Y.~Pei, D.~Morrison, Y.~Tang, and M.~Lewis,
  ``Abstraction of analytical models from cognitive models of human control of
  robotic swarms,'' in {\em Proceedings of the Thirteenth International
  Conference on Cognitive Modeling, Groningen, Germany}, 2015.

\end{thebibliography}

\end{document}